\documentclass[pre,twocolumn,english]{revtex4} 
\usepackage{graphicx}
\usepackage{amsmath,amsthm,amssymb}

\usepackage{bm}

\begin{document}
\title{Information theory and adaptation }

\author{Ilya Nemenman}\email{ilya.nemenman@emory.edu}
\affiliation{Departments of Physics and Biology and
  Computational and Life Sciences Strategic Initiative,
  Emory University, Atlanta, GA 30322}

\begin{abstract} {\em To appear as Chapter 5 of {\em Quantitative
      Biology: From Molecular to Cellular Systems}, ME Wall, ed.
    (Taylor and Francis, 2011).} In this Chapter, we ask questions (1)
  What is the right way to measure the quality of information
  processing in a biological system? and (2) What can real-life
  organisms do in order to improve their
  performance in information-processing tasks? We then review the body
  of work that investigates these questions experimentally,
  computationally, and theoretically in biological domains as diverse
  as cell biology, population biology, and computational neuroscience.
\end{abstract}

\date{\today}
\maketitle

\section{Life is information processing}\label{intro}

All living systems have evolved to perform certain tasks in specific
contexts. There are a lot fewer tasks than there are different
biological solutions that the nature has created. Some of these
problems are universal, while the solutions may be organism-specific.
Thus a lot can be understood about the structure of biological systems
by focusing on understanding of {\em what} they do and {\em why} they
do it, in addtion to {\em how} they do it on molecular or cellular
scales. In particular, this way we can uncover phenomena that
generalize across different organisms, thus increasing the value of
experiments and building a coherent understanding of the underlying
physiological processes.

In this Chapter, we will take this point of view while analyzing what
it takes to do one of the most common, universal functions performed
by organisms at all levels of organization: signal or information
processing and shaping of a response (these are variously known in
different contexts as learning from observations, signal transduction,
regulation, sensing, adaptation, etc.) Studying these types of
phenomena poses a series of well-defined, physical questions: How can
organisms deal with noise, whether extrinsic or generated by intrinsic
stochastic fluctuations within molecular components of information
processing devices? How long should the world be observed before a
certain inference about it can be made? How is the internal
representation of the world made and stored over time? How can
organisms ensure that the information is processed fast enough for the
formed response to be relevant in the ever-changing world? How should
the information processing strategies change when the properties of
the environment surrounding the organism change? In fact, such
``information processing'' questions have been featured prominently in
studies on all scales of biological complexity, from learning
phenomena in animal behavior
\cite{Dayan:2000,Gallistel:2000,Gallistel:2001,Gallistel:2004,Sugrue:2004,Balsam:2009,Balsam:2010},
to analysis of neural computation in small and large animals
\cite{Laughlin:1981,Laughlin:1998,Rieke:1999,Brenner:2000,Reinagel:2000,Fairhall:2001,Liu:2001,Victor:2002,Nemenman:2008},
and to molecular information processing circuits
\cite{Forrest:2000,Arkin:2000,Samoilov:2002,Andrews:2007,Ziv:2007,Tkacik:2008,Tostevin:2009,Celani:2010,Wingreen:2010},
to name just a few.

In what follows, we will not try to embrace the unembraceable, but
will instead focus on just a few questions, fundamental to the study
of signal processing in biology: What is the right way to measure the
quality of information processing in a biological system? and What can
real-life organisms do in order to improve their performance in these
tasks?

\vspace{.1in}
\noindent\framebox[\linewidth][c]
{\noindent\parbox[c]{3.25in}
{\bf The main questions addressed in this review:
\begin{itemize}
\item What is the right way to measure the quality of information
  processing in a biological system? 
\item What can real-life
  organisms do in order to improve their
  performance in these tasks?
\end{itemize}
}
}
\vspace{.1in}

The field of study of biological information processing has undergone
a dramatic growth in the recent years, and it is expanding at an
ever growing rate. There are now entire conferences devoted to the
related phenomena (perhaps the best example is {\em The International
  q-bio Conference on Cellular Information Processing}, \url{http://q-bio.org}, held yearly in Santa Fe,
NM, USA). Hence, in this short chapter, we have neither an ability,
nor a desire to provide an exhaustive literature review. Instead the
reader should keep in mind that the selection of references cited here
is a biased sample of important results in the literature, and I
apologize profusely to my friends and colleagues who find their
deserving work omitted in this overview.
\begin{figure}[b]
\centerline{\includegraphics[width = 6cm]{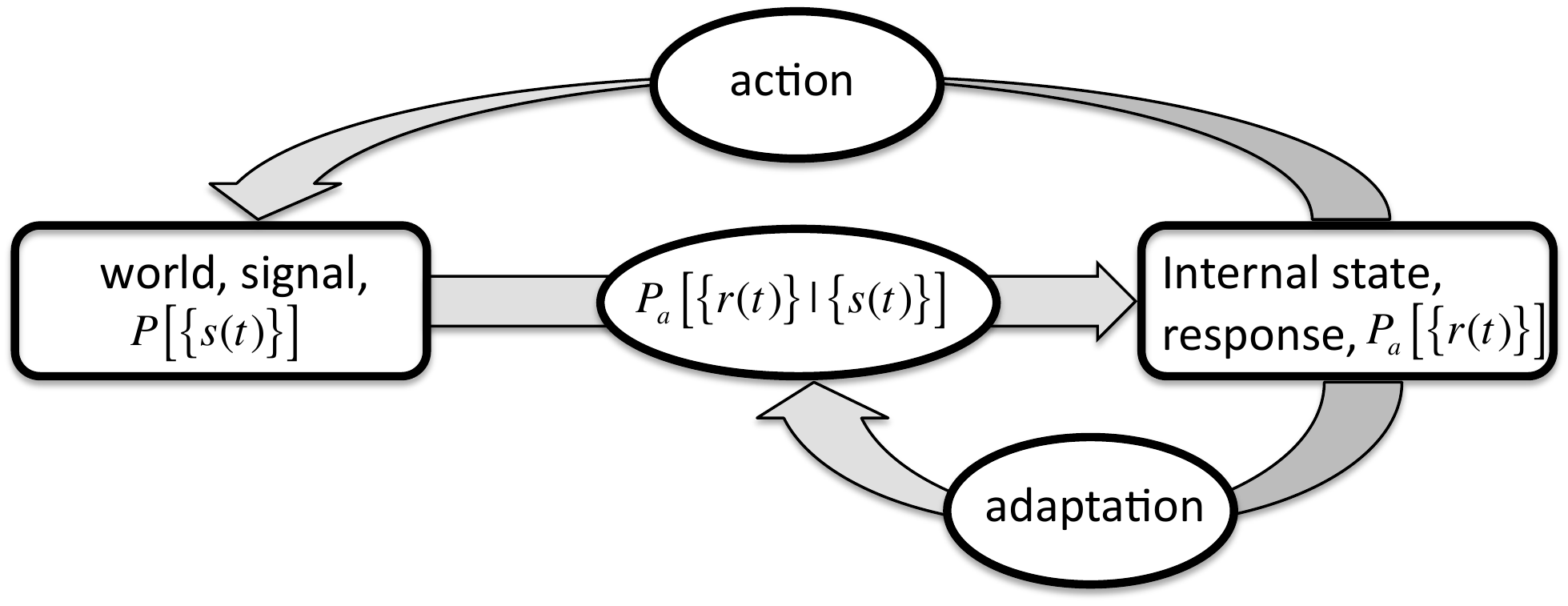}}
\caption{Biological information processing and interactions with the
  world. In this review we leave aside the feedback action between the
  organism internal state and the state of the world and focus on the
  signal processing and the adaptation arrows.}
\label{channel}
\end{figure}

\section{Quantifying biological information processing}

In the most general context, a biological system can be modeled as an
input-output device, cf.~Fig.~\ref{channel} that observes a time-dependent state of the world
$s(t)$ (where $s$ may be intrinsically multidimensional, or even
formally infinite dimensional), processes the information, and
initiates a response $r(t)$ (which can also be very large
dimensional). In some cases, in its turn, the response changes the
state of the world and hence influences the future values of $s(t)$,
making the whole analysis so much harder \cite{Still:2009}. In view of
this, analyzing the information processing means quantifying certain
aspects of the mapping $s(t)\to r(t)$. In this section, we will
discuss the properties that this quantification should possess, and we
will introduce the quantities that satisfy them.

\subsection{What is needed?}
One typically tries to model molecular or other physiological {\em
  mechanisms} of the response generation.  For example, in well-mixed
biochemical kinetics approaches, where $s(t)$ may be a ligand
concentration, and $r(t)$ may be an expression level of a certain
protein, we often write
\begin{equation}
  \frac{dr(t)}{dt}= F_a(r,s,h)-G_a(r,s,h)+\eta_a(r,s,h,t),
\label{main}
\end{equation}
where the nonnegative functions $F_a$ and $G_a$ stand for the
production/degradation of the response, influenced by the level of the
signal $s$, and $\eta$ is a random forcing due to the intrinsic
stochasticity of chemical kinetics at small molecular copy numbers
\cite{Paulsson:2004}. The subscript $a$ stands for the values of
adjustable parameters that define the response (such as various
kinetic rates, concentrations of intermediate enzymes, etc.), which
themselves can change, but on time scales much slower than the
dynamics of $s$ and $r$. In addition, $h$ stands for the activity of
other, hidden cellular state variables, which change according to
their own dynamics, similar to Eq.~(\ref{main}). This dynamics can be
written for many diverse biological information processing systems,
including the neural dynamics, where $r$ will would stand for the
firing rate of a neuron induced by the stimulus \cite{Dayan:2005}.

Importantly, because of the intrinsic stochasticity in
Eq.~(\ref{main}), and because of the effective randomness introduced
by the state of the hidden variables, the mapping between the stimulus
and the response is non-deterministic, and it is summarized in the
probability distribution
$P\left[ \left\{r(t)\right\}|\left\{s(t)\right\}, \left\{h(t)\right\},
  a\right]$, or, marginalizing over $h$,
$P\left[ \left\{r(t)\right\}|\left\{s(t)\right\},
  a\right]\equiv P_a\left[
  \left\{r(t)\right\}|\left\{s(t)\right\}\right]$. In addition, $s(t)$
itself is not deterministic either: other agents, chaotic dynamics,
statistical physics effects, and, at a more microscopic level, even
quantum mechanics conspire to ensure that $s(t)$ can only be specified
probabilistically. Therefore, a simple mapping $s\to r$ is replaced by
a joint probability distribution (note that we will drop the index $a$
in the future where it doesn't cause ambiguities) \begin{multline}
  P\left[ \left\{r(t)\right\}|\left\{s(t)\right\},
  a\right] P[\left\{s(t)\right\}]\\=
P\left[\left\{r(t)\right\},\left\{s(t)\right\}|a\right]\equiv
P_a\left[\left\{r(t)\right\},\left\{s(t)\right\}\right].
\end{multline}
Hence the measure of the quality of the biological information
processing must be a {\em functional} of this joint distribution.

\vspace{.1in}
\noindent\framebox[\linewidth][c]
{\noindent\parbox[c]{3.25in}
{\bf 
Biological information processing is almost always probabilistic.}
}
\vspace{.1in}

Now consider, for example, a classical system studied in cellular
information processing: the {\em E. coli} chemotaxis (see Chapter 15
in this book) \cite{Berg:2004}. This bacterium is capable of swimming
up gradients of various nutrients. In this case, the signal $s(t)$ is
the concentration of such extracellular nutrients. The response of the
system is the activity levels of various internal proteins, like {\em
  cheY}, {\em
  cheA}, {\em cheB}, {\em cheR}, etc., which combine to modulate the
cellular motion through the environment. It is possible to write the
chemical kinetics equations that relate the stimulus to the response
accurately enough and eventually produce the sought after conditional
probability distribution
$P_a[\left\{r(t)\right\}|\left\{s(t)\right\}]$. However, are the
ligand concentrations the variables that the cell ``cares'' about? In
this system, it is reasonable to assume that all protein expression
states that result in the same intake of the catabolite are
functionally equivalent. That is, the goal of the information
processing performed by the cell likely is not to serve as a passive
transducer of the signal into the response
\cite{Celani:2010,Wingreen:2010}, but to make an active computation
that extracts only the part of the signal that is relevant to making
behavioral decisions. We will denote such {\em relevant} aspects of
the world as $e(t)$. For example, for the chemotactic bacterium,
$e(t)$ can be the maximum nutrient intake realizable for a particular
spatiotemporal pattern of the nutrient concentration.

In general, $e$ is not a subset of $s$, or vice versa, and instead the
relation between $s$ and $e$ is also probabilistic,
$P[\left\{e(t)\right\}|\left\{s(t)\right\}]$, and hence the relevant
variable, the signal, and the response form a Markov chain:
\begin{multline}
  P\left[\left\{e(t)\right\}, \left\{s(t)\right\},
  \left\{r(t)\right\}\right]=\\P\left[\left\{e(t)\right\}\right]\, P\left[\left\{s(t)\right\}|
  \left\{e(t)\right\}\right] \,P\left[\left\{r(t)\right\}| \left\{s(t)\right\}\right].
\end{multline}
The quantity we are seeking to characterize the biological information
processing must respect this aspect of the problem. Therefore,
its value must depend explicitly on the choice of the relevance
variable: a computation resulting in the same response will be either
``good'' or ``bad'' depending on what this response is used for. In
other words, one needs to know what the problem is before saying if a solution is good or bad.

\vspace{.1in}
\noindent\framebox[\linewidth][c]
{\noindent\parbox[c]{3.25in}
{\bf 
  It is impossible to quantify the information
    processing without specifying the purpose of the device; that is,
    the relevant quantity that it is supposed to compute.}
}
\vspace{.1in}

\subsection{Introducing the right quantities}
The question of how much can be inferred about a state of a
variable $X$ from measuring a variable $Y$ has been answered by Claude
Shannon over sixty years ago \cite{Shannon:1949}. Starting with basic, uncontroversial
axioms that a measure of information must obey, he derived that the uncertainty in a state of a
variable is given by
\begin{equation}
S[X]=-\sum_xP(x)\log P(x)=-\langle\log P(x)\rangle_{P(x)},
\label{entropy}
\end{equation}
which we know now as the {\em Boltzmann-Shannon entropy}. Here
$\langle\cdots\rangle_{P}$ denotes averaging over the probability
distribution $P$. When the logarithm in Eq.~(\ref{entropy}) is binary
(which we always assume in this Chapter), then the unit of entropy is
a {\em bit}: one bit of uncertainty about a variable means that the
latter can be in one of two states with equal probabilities.

Observing the variable $Y$ (a.k.a.\ {\em conditioning} on it) changes
the probability distribution of $X$, $P(x)\to P(x|y)$, and the
difference between the entropy of $X$ prior to the measurement and the
average conditional entropy tells how informative $Y$ is about $X$:
\begin{align}
I[X;Y]&=S[X]-\langle S[X|Y]\rangle_{P(y)}\\
&=-\left< \log P(x)\right>_{P(x)}+\left<\left< \log
P(y|x)\right>_{P(x|y)}\right>_{P(y)}\\
&=-\left< \log\frac{P(x,y)}{P(x)P(y)}\right>_{P(x,y)}.\label{MI}
\end{align}
The quantity $I[X;Y]$ in Eq.~(\ref{MI}) is known as {\em mutual
  information}. As entropy, it is measured in bits. Mutual information
of one bit means that specifying the variable $Y$ provides us with the
knowledge to answer one yes/no question about $X$.

Entropy and information are additive quantities. That is, when
considering entropic quantities for time series data defined by
$P[\{x(t)\}]$, for $0\le t\le T$, the entropy of the entire series
will diverge linearly with $T$. Therefore, it makes sense to define
entropy and information rates \cite{Shannon:1949} \begin{eqnarray}
  \label{SRate}
  {\mathcal S}[X]&=&\lim_{T\to\infty}\frac{S[x(0\le t<T)\}]}{T},\\
 {\mathcal I}[X;Y]&=&\lim_{T\to\infty}\frac{I[\{x(0\le t<T)\};
  \{y(0\le t<T)\}]}{T},
  \label{IRate}
\end{eqnarray}
which measure the amount of uncertainty in the signal and the
reduction of this uncertainty by the response per unit time.

Entropy and mutual information possess some simple, important
properties \cite{ct:2006}:
\begin{enumerate}
\item Both quantities are non-negative, $0\le S[X]$ and $0\le
  I[X;Y]\le \min \left( S[X],S[Y]\right)$. \item Entropy is zero if
  and only if ({\em iff}) the studied variable is not
  random. Further, mutual information is zero {\em iff}
  $P(x,y)=P(x)P(y)$, that is, there are no any kind of statistical
  dependences between the variables.
\item Mutual information is symmetric, $I[X;Y]=I[Y;X]$.
\item Mutual information is well defined for continuous variables; one
  only needs to replace sums by integrals in Eq.~(\ref{MI}). On the
  contrary, entropy formally diverges for continuous variables (any
  truly
  continuous variable requires infinitely many bits to be specified to
  an arbitrary accuracy), but many properties of entropy are also
  exhibited by the {\em
    differential entropy},
  \begin{equation}
    S[X]=-\int_x dx\, P(x) \log P(x),
  \end{equation}
  which measures the entropy of a continuous distribution relative to
  the uniformly distributed one. In this Chapter, $S[X]$ will always
  mean the differential entropy if $x$ is continuous and the original
  entropy  otherwise.
 \item For a Gaussian
   distribution with a variance of $\sigma^2$,
  \begin{equation}
    S=1/2\log\sigma^2 +{\rm const},
\end{equation}
 and, for a bivariate Gaussian with
  a correlation coefficient of $\rho$,
  \begin{equation}
    \label{mutual_rho}
    I[X;Y]=-1/2\log(1-\rho^2).
\end{equation}
 Thus entropy and mutual information can
  be viewed as generalizations of more familiar notions of variance
  and covariance.
\item Unlike entropy, mutual information is invariant under
  reparameterization of variables. That is
\begin{equation}
I[X;Y]=I[X';Y']
\end{equation}
for all invertible $x'=x'(x),\; y'=y'(y)$. That is, $I$ provides a
measure of statistical dependence between $X$ and $Y$ that is
independent of our subjective choice of the measurement
device\footnote{ From these properties, it is clear that
  mutual information is in some sense a ``nicer'', more fundamental
  quantity. Indeed, even the famous Gibbs paradox in statistical
  physics \cite{Gibbs} is related to the fact that entropy of
  continuous variables is ill-defined. Therefore, it is a pity that
  standard theoretical developments make entropy a primary quantity
  and derive mutual information from it. We believe that it should be
  possible to develop an alternative formulation of information theory
  with mutual information as the primary concept, without introducing
  entropy at all.}.
\end{enumerate}

\subsection{When the relevant variable is unknown: The value of
  information about the world}
One of the most fascinating properties of mutual information is the
Data Processing Inequality \cite{ct:2006}. 

Suppose three variables $X$, $Y$, and $Z$ form a Markov chain,
$P(x,y,z)=P(x)P(y|x)P(z|y)$. In other words, $Z$ is a probabilistic
transformation of $Y$, which, in turn, is a probabilistic
transformation of $X$. Then it can be proven that
\begin{equation}
I[X;Z]\le \min \left( I[X;Y],I[Y;Z]\right).
\end{equation}
That is, {\em you cannot get new information about the original
  variable by further transforming the measured data}; any such
transformation cannot increase the information.

Together with the fact that mutual information is zero {\em iff} the
variables are completely statistically independent, the Data
Processing Inequality suggests that if the variable of interest that
the organism cares about is unknown to the experimenter, then
analyzing the mutual information between the entire input stimulus
(sans noise) and the response may serve as a good proxy. Indeed, due
to the Data Processing Inequality, if $I[S;R]$ is small, then $I[E;R]$
is also small for any mapping $S\to E$ of the signal into the relevant
variable, whether deterministic, $e=e(s)$, or probabilistic, $P(e|s)$.
In many cases, such as \cite{Strong:1998,Ziv:2007,Nemenman:2008}, this
allows us to stop guessing which calculation the organism is trying to
perform and to put an upper bound on the efficiency of the information
transduction, whatever an organism cares about. However, as was
recently shown in the case of chemotaxis in {\em E. coli}, when $e$
and $s$ are substantially different (resource consumption rate vs.\
instantaneous surrounding resource concentration), maximizing $I[S;R]$
is not necessarily what organisms do \cite{Wingreen:2010}.

\vspace{.1in}
\noindent\framebox[\linewidth][c]
{\noindent\parbox[c]{3.25in}
{\bf 
  Information about the outside world is the upper bound on
  information about any of its features.} 
}
\vspace{.1in}

Another reason to study the information about the outside world comes
from the old argument that relates information and game theory
\cite{Kelly:1956}. Namely, consider a zero-sum probabilistic betting
game (think of a roulette without the zeros, where the red an the
black are two equally likely outcomes, and betting on the right
outcome doubles one's investment, while betting on the wrong one leads
to a loss of the bet). Then the logarithmic growth rate of one's
capital is limited from above by the mutual information between the
outcome of the game and the betting strategy. This was recently recast
in the context of population dynamics in fluctuating environments
\cite{Bergstrom:2005,Kussell:2005,Donaldson-Matasci:2008}. Suppose the
environment surrounding a population of genetically identical
organisms fluctuates randomly with no temporal correlations among
multiple states with probabilities $P(s)$. Each organism,
independently of the rest, may choose among a variety of phenotypical
decisions $d$, and the log-growth rate depends on the pairing of $s$
and $d$. Evolution is supposed to maximize this rate, averaged over
long times. However, the current state of the environment is not
directly known, and the organisms may need to respond
probabilistically. While the short-term gain would suggest choosing
the response that has the highest growth rate for the most probable
environment, the longer term strategy would require bet-hedging
\cite{Seger:1987}, with different individuals making different
decisions.

Suppose an individual now observes the environment and gets an
imperfect internal representation of it, $r$, with the conditional
probability of $P(r|s)$. What is the value of this information? Under
very general conditions, this information about the environment can
improve the log-growth rate by as much as $I[S;R]$
\cite{Bergstrom:2005}. In more general scenarios, the maximum
log-growth advantage over uninformed peers needs to be discounted by
the cost of obtaining the information, by the delay in getting it
\cite{Kussell:2005}, and, more trivially, by the ability of the
organism to utilize it. Therefore, while these brief arguments are far
from painting a complete picture of relation between information and
natural selection, it is already clear that maximization of the
information between the surrounding world and the internal response to
it is not an academic exercise, but is directly related to fitness and
will be selected for by evolution.\footnote{Interestingly,
  it was recently argued \cite{Frank:2009} that natural selection,
  indeed, serves to maximize the
  information that a population has about its environment, providing
  yet another evidence for the importance of information-theoretic
  considerations in biology.}

\vspace{.1in}
\noindent\framebox[\linewidth][c]
{\noindent\parbox[c]{3.25in}
{\bf
Information about the outside world puts an upper bound on the
  fitness advantage of an individual over uninformed peers.} 
}
\vspace{.1in}

It is now well known that probabilistic bet hedging is the strategy
taken by bacteria for survival in the presence of antibiotics
\cite{Balaban:2004,Kussell:2005g} and for genetic recombination
\cite{Maamar:2007,Cagatay:2009,Wylie:2010}. In both cases, cell
division (and hence population growth) must be stopped either to avoid
DNA damage by antibiotics, or to incorporate newly acquired DNA into
the chromosome. Still, a small fraction of the cells choose not to
divide even in the absence of antibiotics to reap the much larger
benefits if the environment turns sour (these are called the {\em
  persistent} and the DNA uptake {\em competent} bacteria for the two
cases, respectively). However, it remains to be seen in an experiment
if real bacteria can reach the maximum growth advantage allowed by the
information-theoretic considerations. Another interesting possibility
is that cancer stem cells and mature cancer cells also are two
probabilistic states chosen to hedge bets against interventions of
immune systems, drugs, and other surveillance mechanisms
\cite{Bowen:2009}.

\subsection{Time dependent signals: Information and Prediction}
In many situations, such as persistence in the face of antibiotics
treatment mentioned above, an organism settles into a certain
response much faster than the environment has a chance to change
again. In these cases, it is sufficient to consider the same-time
mutual information between the signals and the responses, as in
\cite{Ziv:2007}, $I[s(t);r(t)]=I[S;R]$, which is what we've been doing
up to now.

More generally, formation of any response takes time, which may be
comparable to time scales of changes of the stimuli. What are the
relevant quantities to characterize biological information processing
in such situations? Traditionally, one either considers delayed
informations \cite{Arkin:1995},
\begin{equation}
I_\tau[S;R]=I[s(t);r(t+\tau)],
\end{equation}
where $\tau$ may be chosen as $\tau=\arg\max_{t'} I[s(t);r(t+t')]$, or
studies information rates, as in Eq.~(\ref{IRate}). The first choice
measures the information between the stimulus and the response most
constrained by it; typically this would be the response formed a
certain characteristic signal transduction time after the stimulus
occurrence. The second approach looks at correlations between all
possible pairs of stimuli and responses separated by different delays.

While there are plenty of examples of biological systems where one or
the other of these quantities is optimized, both of these approaches
are insufficient. ${\mathcal
  I}_\tau$ doesn't represent all of the gathered information since
bits at different moments of time are not independent of each other.
Further, it does not take into the account that temporal correlations
in the stimulus allow to predict it, and hence the response may be
formed even before the stimulus occurs. On the other hand, the
information rate does not distinguish among predicting the signal,
knowing it soon after it happens, or having to wait for $T\to\infty$
in order to be able to estimate it from the response.

To avoid these pitfalls, one can consider the information available to
an organism that is relevant for specifying not all of the stimulus,
but only of its future. Namely, we can define the {\em predictive
  information} about the stimulus available from observation of a
response to it of a duration $T$, \begin{equation}
   I_{\rm pred}[R(T);S]=I[ \{r(-T\le t\le0)\};\{s(t>0)\}].
\end{equation}
This definition is a generalization of the one used in
\cite{Bialek:2001}, which had $r(t)=s(t)$, and hence calculated the
upper bound on $I_{\rm pred}$ over all possible sensory schemes
$P[\{r(t)\}|\{s(t)\}]$. 

All of the $I_{\rm
  pred}$ bits are available to be used instantaneously, and there is
no specific delay $\tau$ chosen a priori and semi-arbitrarily. The
predictive information is nonzero only to the extent that the signal
is temporally correlated, and hence the response to its past values
can say something about its future. Thus focusing on predictability
may resolve a traditional criticism of information theory that bits
don't have an intrinsic meaning and value, and some are more useful
than the others: since any action takes time, only those bits have
value that can be used to predict the stimulus at the time of action,
that is, in the future \cite{Bialek:2001,Creutzig:2009}.

\vspace{.1in}
\noindent\framebox[\linewidth][c]
{\noindent\parbox[c]{3.25in}
{\bf
Predictive information allows to assign an objective value
  to information: only those bits are useful that can be used to guide
  future responses.}
}
\vspace{.1in}

The notion of predictive information is conceptually appealing, and
there is clear experimental and computational evidence that behavior
of biological systems, from bacteria to mammals, is consistent with
attempting to make better predictions (see, for example,
\cite{Srinivasan:1982,Schwartz:2007,Nemenman:2008,Hosoya:2005,Vergassola:2007,Tagkopoulos:2008,Mitchell:2009}
for just some results). However, even almost ten years after $I_{\rm
  pred}$ was first introduced, it still remains to be seen
experimentally if optimizing predictive information is one of the
objectives of biological systems, and whether population growth rates
in temporally correlated environments can be related to the amount of
information available to predict them. Some of the reasons for the
relative lack of progress may be practical considerations that
estimation of informations among nonlinearly related multidimensional
variables \cite{Nemenman:2002,Paninski:2003,Nemenman:2008} or
extracting the predictive aspects of the information
\cite{Tishby:1999} from empirical data is hard, while for simple
Gaussian signals and responses with finite correlation times,
optimization of predictive information reduces to a much more prosaic
matching of Wiener extrapolation filters \cite{Wiener:1964}.

\section{Improving Information-Processing Performance}

Understanding the importance of information about the outside world
and knowing which quantities can be used to measure it, we are faced
with the next question: How can the available information be increased
in view of the limitations imposed by the physics of the signal and of
the processing device, such as stochasticity of molecular numbers and
arrival times, or energy constraints?

\subsection{Strategies for Improving The Performance}
We start with three main theorems of information theory due to Shannon
\cite{Shannon:1949}. In the {\em source coding theorem}, he proved
that to record a signal without losses, one needs only ${\cal S}$, the
signal entropy rate, bits per unit time. In the {\em
  channel coding theorem}, he showed that the maximum rate of errorless
transmission of information through a channel specified by
$P[\{r(t)\}|\{s(t)\}]$ is given by $C=\max_{P(\{s(t)\}}{\cal I}[R;S]$,
which is called the channel capacity. Finally, the {\em rate
  distortion theorem} calculates the minimum size of the message that
must be sent error-free in order to recover the signal with an
appropriate mean level of some pre-specified distortion measure. None
of these theorems considers the time delay before the message can be
decoded, and typically one would need to wait for very long times and
accumulate long message sequences to reach the bounds predicted by the
theorems since, for example, responses long time away from a certain
signal may still carry some information about it.

Leaving aside the complication of dynamics, which one may hope to
solve some day using the predictive information ideas, these theorems
tell us exactly what an organism can do to optimize the amount of
information it has about the outside world. First, one needs to
compress the measured signal, removing as many redundancies as
possible. There is evidence that this happens in a variety of
signaling systems, starting with the classical
Refs.~\cite{Barlow:1959,Barlow:1961,Srinivasan:1982,Atick:1990}.
Second, one needs to encode the signal in a way that allows the
transmitted information to approach the channel capacity limit by
remapping different values of signal into an intermediate signaling
variable whose states are easier to transmit without an error. Again,
there are indications that this happens in living systems
\cite{Marshall:1995,Sabbagh:2001,Lahav:2004,Ma:2005,Hensen:2008,Tkacik:2008}.
Finally, one may choose to focus only on important aspects of the
signal, such as communicating changes in the signal, or thresholding
its value \cite{Huang:1996,Markevich:2004,Berg:2004,Nemenman:2008}.

If the references in the previous paragraph look somewhat thin, it is
by choice since neither of these approaches are unique to biology,
and, in fact, most artificial communication system use them: a cell
phone filters out audio frequencies that are irrelevant to human
speech, compresses the data, and then encodes it for sending with the
smallest possible errors. A lot of engineering literature discusses
these questions \cite{ct:2006}, and we will not touch them here
anymore. What makes biological systems unique is an ability to improve
the information transmission by modifying their own properties in the
course of their life. This adjusts the $a$ in
$P_a[\{r(t)\}|\{s(t)\}]$, and hence modifies the conditional
probability distribution itself. This would be equivalent to a cell
phone being able to change its physical characteristics on the fly.
Unfortunately, as the recent issues with the iPhone antenna have
shown, human engineered systems are no match to biology in this
regard: they are largely incapable of adjusting their own design if
the original turns out to be flawed.

\vspace{.1in}
\noindent\framebox[\linewidth][c]
{\noindent\parbox[c]{3.25in}
{\bf 
  Unlike most artificial systems, living organisms can change
    their own properties to
    optimize their information processing.}
}
\vspace{.1in}

The property of changing one's own characteristics in response to the
observed properties of the world is called {\em adaptation}, and the
remainder of this section will be devoted to its overview. In
principle, we make no distinction whether this adaptation is achieved
by natural selection or by physiological processes that act on much
faster times scales (comparable to the typical signal dynamics), and
sometimes the latter may be as powerful as the former
\cite{Ziv:2007,Mugler:2009}. Further, we note that adaptation of the
response probability distribution and formation of the response itself
are, in principle, a single process of formation of the response on
multiple time scales. Our ability to separate it into a fast response
and a slow adaptation (and hence much of the discussion below) depends
on existence of two well-separated time scales in the signal and in
the mechanism of the response formation. While such clear separation
is possible in some cases, it is harder in others, and especially when
the time scales of the signal and the fast response may be changing
themselves. Cases without a clear separation of scales raise a variety
of interesting questions, but we will leave them aside for this
discussion.

\subsection{Three Kinds of Adaptation in Information Processing}

We often can linearize the dynamics,
Eq.~(\ref{main}), to get the following equation describing formation
of small responses\begin{equation}
  \frac{dr}{dt}=f\left[s\left(t\right)\right]-kr +\eta(t,r,s).
\label{filter}
\end{equation}
Here $r$ may be an expression of an mRNA following activation by a
transcription factor $s$, or the firing rate of a neuron following
stimulation. In the above expression, $f$ is the response activation
function, which depends on the current value of the signal; $k$ is
the rate of the first-order relaxation or degradation; and $\eta$ is
some stochastic process representing the intrinsic noisiness of the
system. In this case, $r(t)$ depends on the entire history of $\{s(t')\}$,
$t'<t$, and hence carries some information about it as well. 

\begin{figure}[t]
\centerline{\includegraphics[width = 8cm]{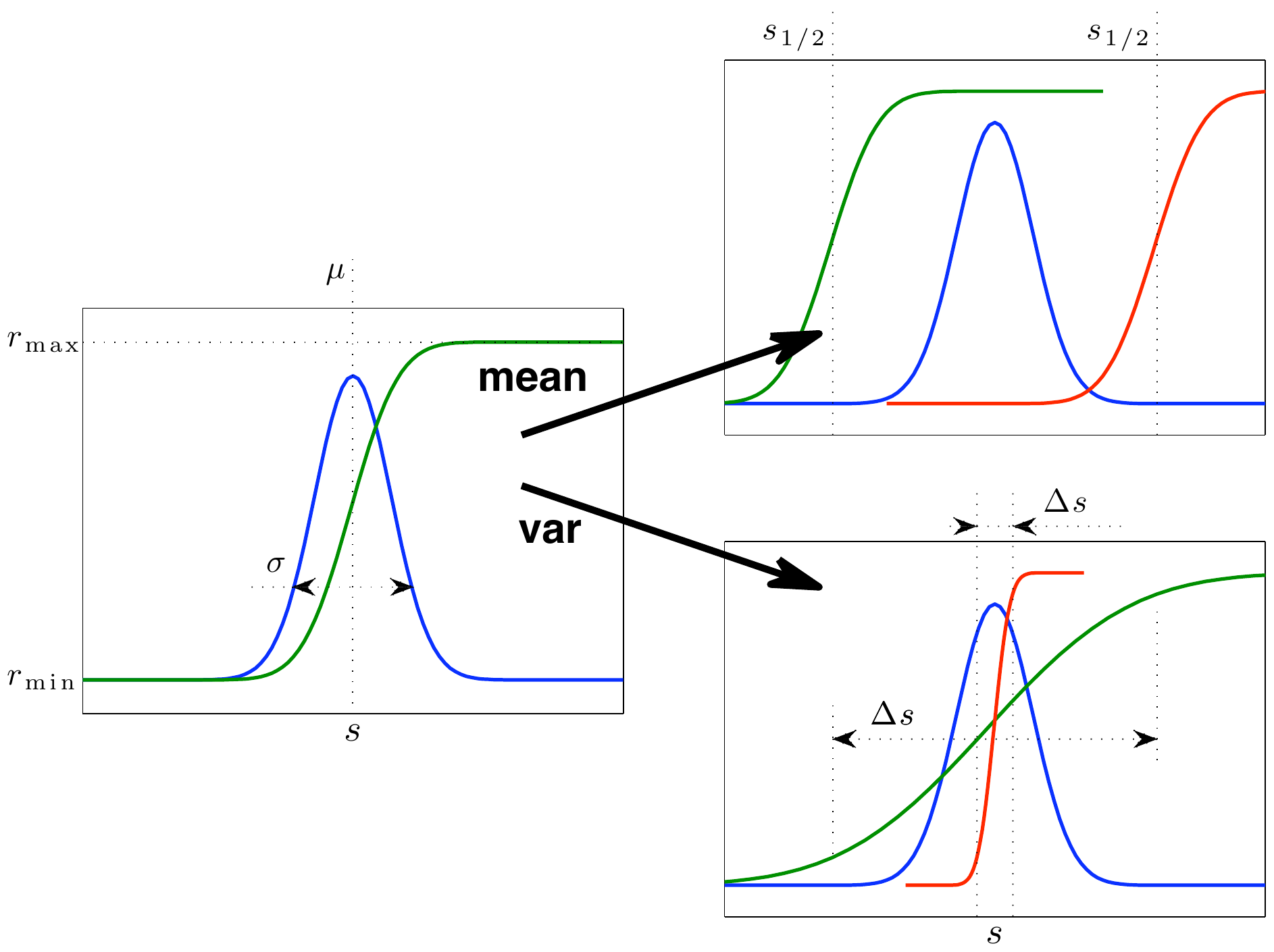}}
\caption{Parameters characterizing response to a signal. Left panel:
  the probability distribution of the signal, $P(s)$ (blue), and the
  best-matched steady state dose-response curve $r_{\rm ss}$
  (green). Top right: mismatched response midpoint. Bottom right:
  mismatched response gain.}
\label{response}
\end{figure}

For quasi-stationary signals (that is, the correlation time of the
signal, $\tau_s\gg1/k$), we can write the steady state dose-response (or
firing rate, or \dots) curve 
\begin{equation}
  r_{\rm ss}=f\left[s(t)\right]/k,
\label{dose-response}
\end{equation}
and this will be smeared by the noise $\eta$. 
A typical monotonic sigmoidal $f$ is characterized by only a few
large-scale parameters: the range, $f_{\min}$ and $f_{\max}$; the
argument $s_{1/2}$ at the mid-point value $(f_{\min}+f_{\max})/2$; 
and the width of the transition region, $\Delta
s$ (see Fig.~\ref{response}). If the mean signal $\mu\equiv\langle
s(t)\rangle_t\gg s_{1/2}$, then, for most signals, $r_{\rm ss}\approx
f_{\rm max}/k$ and responses to two typical different signals $s_1$
and $s_2$ are indistinguishable as long as
\begin{equation}
\left.\frac{dr_{\rm ss}(s)}{ds}\right|_{s=s_1}(s_2-s_1)<\sigma_\eta/k,
\end{equation}
where $\sigma_\eta/k$ is the precision of the response resolution
expressed through the standard deviation of the noise. Similar
situation happens when $\mu\ll s_{1/2}$ and $r_{\rm ss}\approx f_{\rm
  min}/k$. Thus, to reliably communicate information about the signal,
$f$ should be tuned such that $s_{1/2} \approx \mu$. If a real
biological system can perform this adjustment, we call this {\em
  adaption to the mean} of the signal, {\em desensetization}, or {\em
  adaptation of the first kind}. If $s_{1/2}(\mu)=\mu$, then the
adaptation is {\em perfect}. This kind of adaptation has been observed
experimentally and predicted computationally in a lot more systems
than we can list here, including phototransduction, neural and
molecular sensing, multistate receptor systems, immune response, and
so on, with active work persisting to date (see, e.g.,
Refs.~\cite{Iglesias:2003,Berg:2004,Norman:1979,MacGlashan:1998,Detwiler:2000,Rao:2004,Hensen:2008,Friedlander:2009,Muzzey:2009,Anishkin:2009,Belyy:2010}
for a very incomplete list of references on the subject).
For example, the best studied adaptive circuit in molecular biology,
the control of chemotaxis of {\em E.\
  coli} (see Chapter 15), largely produces adaptation of the first
kind \cite{Alon:1999,Hansen:2008}. Further, a variety of problems in
synthetic biology are due precisely to the mismatch between the
typical protein concentration of the input signal and the response
function that maps this concentration into the rate of mRNA
transcription or protein translation (cf.~\cite{Salis:2009} and
Chapter 4 in this book). Thus there is an active community of
researchers working on endowing these circuits with proper adaptive
matching abilities of the first kind.

Consider now the quasi-stationary signal taken from the distribution
with $\sigma\equiv \left(\langle
  s(t)^2\rangle _t-\mu^2\right)^{1/2}\gg \Delta s$. Then the response
to most of the signals is indistinguishable from the extremes, and it
will be near the midpoint $\sim (r_{\max}+r_{\min})/2$ if $\sigma\ll
\Delta s$. Thus, to use the full dynamic range of the response, a
biological system must tune the width of the sigmoidal dose-response
curve to $\Delta s\approx\sigma$. We call this {\em
  gain control}, {\em variance adaptation}, or {\em adaptation of
  the second kind}. Experiments show that a variety of systems exhibit
this adaptive behavior as well \cite{Endres:2008}, especially in the
context of neurobiology \cite{Brenner:2000,Borst:2005}, and maybe even
of evolution \cite{Kashtan:2005}. 

These matching strategies are well known in signal processing
literature under the name of histogram equalization. Surprisingly,
they are nothing but a special case of optimizing the mutual
information $I[S;R]$, as has been shown first in the context of
information processing in fly photoreceptors \cite{Laughlin:1981}.
Indeed, for quasi-steady state responses, when noises are small
compared to the range of the response, the arrangement that optimizes
$I[S;R]$ is the one that produces $P(r)\propto1/\sigma_{r|s}$. In
particular, when $\sigma_\eta$ is independent of $r$ and $s$, this
means that each $r$ must be used equiprobably, that is,
$f^*(s)=\int_{-\infty}^sP(s')ds'$. Adaptation of the first and the
second kind follows from these considerations immediately. In more
complex cases, when the noise variance is not small or not constant,
derivation of the optimal response activation function cannot be done
analytically, but numerical approaches can be used instead. In
particular, in transcriptional regulation of the early {\em
  Drosophila} embryonic development, the matching between the response
function and the signal probability distribution has been observed for
nonconstant $\sigma_{r|s}$ \cite{Tkacik:2008}. However, we caution the
reader that, even though adaptation {\em can} have this intimate
connection to information maximization, and it is essentially
omni-present, the number of systems where the adaptive strategy has
been analyzed quantitatively to show that it results in optimal
information processing is not that large.

We now relax the requirement of quasi-stationarity and return to
dynamically changing stimuli. We rewrite Eq.~(\ref{filter}) in the
frequency domain, \begin{equation}
 r_\omega =
  \frac{[f(s)]_\omega +\eta_\omega}{k+i\omega}, \label{filterw} 
\end{equation}
which shows that the simple first order (or linearized) kinetics
performs low pass filtering of the nonlinearly transformed signal
\cite{Arkin:2000,Samoilov:2002}. As discussed long ago by Wiener
\cite{Wiener:1964}, for given temporal correlations of the stimulus
and the noise (which we summarize here for simplicity by correlation
times $\tau_s$ and $\tau_\eta$), there is an optimal cutoff frequency
$k$ that allows to filter out as much noise as possible without
filtering out the signal. Change of the parameter $k$ to match the
temporal structure of the problem is called the {\em time scale
  adaptation} or {\em adaptation of the third
  kind}. Just like the first two kinds, time scale adaptation also can
be related to maximization of the stimulus-response mutual information
by means of a simple observation that minimization of the quadratic
prediction error of the Wiener filter is, under certain assumptions,
equivalent to maximizing information about the signal,
cf.~Eq.~(\ref{mutual_rho}). 

This adaptation strategy is difficult to study experimentally since
(a) detection of variation of the integration cutoff frequency $k$
potentially requires observing the adaptation dynamics on very long
time scales, and (b) prediction of optimal cutoff frequency requires
knowing the temporal correlation properties of signals, which are far
from trivial to measure (see, e.g., Ref.~\cite{Reinagel:2001} for a
review on literature on analysis of statistical properties of natural
signals). Nonetheless, experimental systems as diverse as turtle cones
\cite{Baylor:1974}, rats in matching foraging experiments
\cite{Gallistel:2001}, mice retinal ganglion cells \cite{Wark:2009},
and barn owls adjusting auditory and visual maps \cite{Knudsen:2002}
show adaptation of the filtering cutoff frequency in response to
changes in the relative time scales and/or the variances of the signal
and the noise. In a few rare cases, including fly self-motion
estimation \cite{Fairhall:2001} and {\em E.\
  coli} chemotaxis \cite{Andrews:2006} (numerical experiment), it
turned out to be possible to show that the time scale matching not
only improves, but optimizes the information transmission.

\vspace{.1in}
\noindent\framebox[\linewidth][c]
{\noindent\parbox[c]{3.25in}
{\bf The three kinds of adaptation (to the mean, to the variance,
    and to the time scale of change of the signal) can all be related to
    maximization of the stimulus-response information.}
}
\vspace{.1in}

Typically one considers adaptation as a phenomenon different from
redundancy reduction, and we have accepted this view. However, there
is a clear relation between the two mechanisms. For example,
adaptation of the first kind can be viewed as subtracting out the mean
of the signal, stopping its repeated, redundant transmission and
allowing to focus on the non-redundant, changing components of the
signal. As any redundancy reduction procedure, this may introduce
ambiguities: a perfectly adapting system will respond in the same
fashion to different stimuli, preventing unambiguous identification of
the stimulus based on the instantaneous response. Knowing statistics
of responses on the scale of adaptation itself may be required to
resolve the problem. This interesting complication has been explored
in a few model systems \cite{Fairhall:2001,Wark:2009}.

\subsection{Mechanisms of Different Adaptations}
The three kinds of adaptation we consider here can all be derived from
the same principle of optimizing the stimulus-response mutual
information, and evolution can achieve all of them. However, the
mechanisms behind these adaptations on physiological, non evolutionary
time scales and their mathematical descriptions can be substantially
different, as we describe below.

\begin{figure}[t]
\centerline{\includegraphics[width = 8cm]{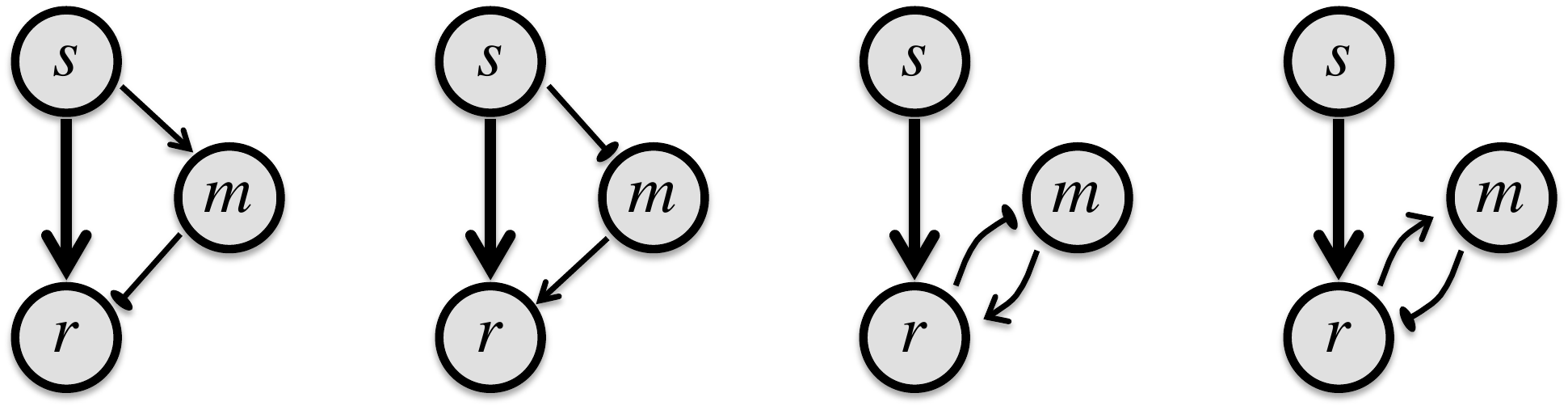}}
\caption{Different network topologies able to perform adaptation to
  the mean. The nodes are labeled: $s$ - signal, $r$ - response, and
  $m$ - memory. Sharp arrows indicate activation/excitation and blunt ones stand
  for deactivation/suppression. The thickness of arrows denotes the
  speed of action (faster action for thicker arrows).} \label{loops}
\end{figure}

The adaptation of the first kind has been studied extensively. On
physiological scales, it is implemented typically using negative
feedback loops or incoherent feedforward loops, as illustrated in
Fig.~\ref{loops}. In all of these cases, the fast activation of the
response by the signal is then followed by a delayed suppression
mediated by a memory node. This allows the system to transmit changes
in the signal, and yet to desensetize and return close (and sometimes
perfectly close) to the original state if the same excitation persist.
This response to {\em changes} in the signal earns adaptation of the
first kind the name of {\em
  differentiating filter}. In particular, the feedback loop in {\em
  E.\ coli} chemotaxis \cite{Berg:2004,Alon:1999} or yeast signaling
\cite{Yu:2008} can be represented as the feedback topologies in the
figure (see Chapter 15), and different models of {\em Dictyostelium}
adaptation include both feedforward and feedback designs
\cite{Iglesias:2003,Yang:2006}.

 The different network topologies have different
sensitivities to changes in the internal parameters, different
tradeoffs between the sensitivity to the stimulus change and the
quality of adaptation, and so on. However, fundamentally they are
similar to each other. This can be seen by noting that since the goal
of these adaptive system is to keep the signal within the {\em small}
transition region between the minimum and the maximum activation of
the response, it makes sense to linearize the dynamics of the networks
near the mean values of the signal and the corresponding response.
Defining $\xi=s-\bar{s}$, $\zeta=r-\bar{r}$, and $\chi=m-\bar{m}$, one
can write, for example, for the the feedback topologies in
Fig.~\ref{loops} \begin{eqnarray}
  \frac{d \zeta}{dt}&=& -k_{\zeta\zeta}\zeta
  +k_{\zeta\xi}\xi-k_{\zeta\chi}\chi+\eta_\zeta.
\label{fb1}\\
  \frac{d \chi}{dt}&=& k_{\chi\zeta}\zeta
  -k_{\chi\chi}\chi+\eta_\chi,
\label{fb2}
\end{eqnarray}
where $\eta_{\cdot}$ are noises, and the coefficients $k_{**}$
are positive for the fourth topology, and some of them change their
signs for the third. Doing the usual Fourier transform of these
equations (see Ref.~\cite{Detwiler:2000} for a very clear, pedagogical
treatment) and expressing $\zeta$ in terms of $\xi$, $\eta_\zeta$, and
$\eta_\chi$, we see that it is only the product of
$k_{\chi\zeta}k_{\zeta\chi}$ that matters for the properties of the
filter, Eq.~(\ref{fb1}, \ref{fb2}). Hence both the feedback topologies
in Fig.~\ref{loops} are essentially equivalent in this regime.
Furthermore, as argued in \cite{Iglesias:2003,Sontag:2010}, a simple
linear transformation of $\zeta$ and $\chi$ allows to recast the
incoherent feedforward loops (the two first topologies in
Fig.~\ref{loops}) into a feedback design, again arguing that, at least
in the linear regime, the differences among all of these organizations
are rather small from the mathematical point of view.\footnote{Sontag
  has recently
  considered the case where the linear term in the feedforward and/or
  feedback interactions is zero, and the leading coupling term in the
  dynamics of $\zeta$ is bilinear; there
  the distinctions among the topologies are somewhat more tangible
  \cite{Sontag:2010}.}

The reason why we can make so much progress in the analysis of
adaptation to the mean is that the mean is a linear function of the
signal, and hence it can be accounted for in a linear approximation.
Different network topologies differ in their actuation components
(that is, how the measured mean is then fed back into changing the
response generation), but averaging a linear function of the signal
over a certain time scale is the common description of the sensing
component of essentially all adaptive mechanisms of the first kind.

\vspace{.1in}
\noindent\framebox[\linewidth][c]
{\noindent\parbox[c]{3.25in}
{\bf Adaptation to the mean can be analyzed linearly, and many
    different designs become similar in this regime.} 
}
\vspace{.1in}

Variance and time scale adaptations are fundamentally different. While
the actuation part for them is not any more difficult than for
adaptation to the mean, adapting to the variance requires averaging
the square or another nonlinear function of the signal to sense its
current variance, and estimation of the time scale of the signal
requires estimation of the spectrum or of the correlation function
(both are bilinear averages). Therefore, maybe it is not surprising
that the literature on mathematical modeling of mechanisms of these
types of adaptation is rather scarce. While functional models
corresponding to a bank of filters or estimators of environmental
parameters operating at different time scales can account for most of
the experimentally observed data about changes in the gain and in the
scale of temporal integration
\cite{DeWeese:1998,Brenner:2000,Fairhall:2001,Gallistel:2001,Wark:2009},
to our knowledge, these models largely have not been related to
non-evolutionary, mechanistic processes at molecular and cellular
scales that underlie them.

The largest inroads in this direction have been achieved when
integration of a nonlinear function of a signal results in an
adaptation response that depends not just on the mean, but also on
higher order cumulants of the signal, effectively mixing different
kinds of adaptation together. This may be desirable in the cases of
photoreception \cite{Detwiler:2000} and chemosensing
\cite{Endres:2008}, where the signal mean is unalieanbly connected to
the signal or the noise variances (e.g, the standard deviation of
brightness of a visual scene scales linearly with the background
illumination, while the noise in the molecular concentration is
proportional to the square root of the latter). Similarly, mixing
means and variances allows the budding yeast to respond to {\em
  fractional} rather than additive changes of a pheromone
concentration \cite{Paliwal:2007}. In other situations, like
adaptation by a receptor with state-dependent inactivation properties,
similar mixing of the mean signal with its temporal correlation
properties to form an adaptive response may not serve an obvious
purpose \cite{Friedlander:2009}.

\vspace{.1in}
\noindent\framebox[\linewidth][c]
{\noindent\parbox[c]{3.25in}
{\bf We know very little about physiological mechanisms of
    adaptation of the second and the third kind.} 
}\vspace{.1in}

In a similar manner, integration of a strongly nonlinear function of a
signal may allow a system to respond to signals in a gain-insensitive
fashion, effectively adapting to the variance without a true
adaptation. Specifically, one can threshold the stimulus around its
mean value and then integrate it to count how long it has remained
positive. For any temporally correlated stimulus, the time since the
last mean-value crossing is correlated to the instantaneous stimulus
value (it takes long time to reach high stimulus values), and this
correlation is independent of the gain. It has been argued that
adaptation to the variance in fly motion estimation can be explained
at least in part by this non-adaptive process \cite{Borst:2005}.
Similar mechanisms are easy to implement in molecular signaling
systems as well \cite{Nemenman:2011}.

\section{What's next?}
It is clear beyond that information theory has an important role in
biology. It is a mathematically correct construction for analysis of
signal processing systems. It provides a general framework to recast
adaptive processes on scales from evolutionary to physiological in
terms of a (constrained) optimization problem. Sometimes it even makes
(correct!) predictions about responses of living systems following
exposure to various signals.

So, what's next for information theory in the study of signal
processing in living systems?

The first, and the most important problem that still remains to be
solved is that many of the stories we mentioned above are incomplete.
Since we never know for sure which specific aspect of the world,
$e(t)$, an organism cares about, and the statistics of signals are
hard to measure in the real world, an adaptation that seems to
optimize $I[S;R]$ may be an artifact of our choice of $S$ and of
assumptions about $P(s)$, but not a consequence of the quest for
optimality by an organism. For example, the time scale of filtering in
{\em E.\
  coli} chemotaxis \cite{Andrews:2006} may be driven by the
information optimization, or it may be a function of very different
pressures. Similarly, a few standard deviations mismatch between the
cumulative distribution of light intensities and a photoreceptor
response curve in fly vision \cite{Laughlin:1981} can be a sign of an
imperfect experiment, or it can mean that we simply got (almost)
lucky, and the two curves nearly matched by chance. It is difficult to
make conclusions based on one data point!

Therefore, to complete these and similar stories, the information
arguments must be used to make predictions about adaptations in novel
environments, and such adaptations must be observed experimentally.
This has been done in some contexts in neuroscience
\cite{Gallistel:2000,Brenner:2000,Fairhall:2001,Witten:2008}, but
molecular sensing lags behind. This is largely because evolutionary
adaptation, too slow to observe, is expected to play a major role
here, and because careful control of dynamic environments, or
characterization of statistical properties of naturally occuring
environments \cite{Reinagel:2001} needed for such experiments is not
easy. New experimental techniques, such as microfluidics
\cite{Melin:2007} and artificially sped up evolution
\cite{Poelwijk:2007} are about to solve these problems, opening the
proverbial doors wide open for a new class of experiments.

The second important research direction, which will require combined
progress in experimental techniques and mathematical foundations, is
likely going to be the return of dynamics. This has had a
revolutionary effect in neuroscience \cite{Rieke:1999}, revealing
responses unimaginable for quasi-steady-state stimuli, and dynamical
stimulation is starting to take off in molecular systems as well
\cite{Mettetal:2008,Hensen:2008}. How good are living systems in filtering out
those aspects of their time-dependent signals that are not predictive
and are, therefore, of no use? What is the evolutionary growth bound
when signals change in a continuous, predictive fashion? None of these
questions have been touched yet, whether theoretically or experimentally.

Finally, we need to start building mechanistic models of adaption in
living systems that are more complex than a simple subtraction of the
mean. How are the amazing adaptive behaviors of the second and the
third kind achieved in practice on physiological scales? Does it even
make sense to distinguish the three different adaptations, or can some
molecular or neural circuits achieve them all? How many and which
parameters of the signal do neural and molecular circuits estimate and
how? Some of these questions may be answered if one is capable of
probing the subjects with high frequency, controlled signals
\cite{Nemenman:2005}, and the recent technological advances will be a
gamechanger as well.

Overall, studying biological information processing over the next ten
years will be an exciting pastime!

\begin{acknowledgements}
  I am grateful to Michael E. Wall for asking me to write this book
  chapter. I would like to thank Sorin Tanase Nicola, and
  H.~G.~E.~Hentschel for insightful comments about the manuscript.
\end{acknowledgements}

\end{document}